%% file: iono_chapter.tex
\documentclass[a4paper,11pt]{article}
\usepackage{aaskaiid}
\usepackage{xcolor}
\usepackage[english]{babel}
\usepackage{epsfig}
\usepackage{epstopdf}
\usepackage{psfrag}
\usepackage{color}
\usepackage{graphicx}
\usepackage{fancyhdr}
\usepackage{graphics} 
\usepackage{soul}
\usepackage[normalem]{ulem}
\usepackage{enumitem}
\usepackage{bm}
\usepackage{hyperref}
\usepackage{subcaption}
\usepackage{comment}
\usepackage[utf8]{inputenc}
\usepackage{newunicodechar}
\DeclareUnicodeCharacter{2212}{\textminus}
\usepackage{amsmath}
\usepackage{xspace}
\usepackage{xcolor}
\usepackage{orcidlink}

\graphicspath{{Figures/}} 

\newcommand{\dtec}{$\delta\rm{TEC}$\xspace}
\newcommand{\hi}{H{\sc i}\xspace}

\title{Studying Ionosphere Using SKA-Low and SKA-Mid}
\ShortTitle{Ionosphere with the SKA}
\input{authors}

\ShortName{Datta et al.} 

\abstract{The Earth's ionosphere introduces systematic effects that limit the performance of radio interferometers operating at low frequencies ($\lesssim 1$\,GHz). These ionospheric effects intensify during periods of heightened geomagnetic activity or for observations with extended baseline configurations. As each Pathfinder telescope operates at a different magnetic latitude, they experience distinct ionospheric regimes, offering complementary insights into ionospheric behaviour. In this work, we present a comparative study of ionospheric disturbances using observations from the uGMRT, VLA, MWA, and LOFAR, spanning a wide range of geographic and geomagnetic conditions. We present both antenna-based and field-based analyses to quantify phase fluctuations, positional offsets, and scintillation effects across these arrays. The measured total electron content (TEC) gradients reveal variations in spatial and temporal ionospheric structures with sensitivities that exceed those achievable with Global Navigation Satellite System (GNSS) measurements. By combining multi-telescope results, we assess the impact of ionospheric turbulence on calibration and imaging fidelity, and use these findings to forecast the expected ionospheric effects on observations with SKA-Low and SKA-Mid.}

\begin{document}
\maketitle


\section{Introduction}\label{sec:introduction}
Radio signals arriving from beyond Earth’s atmosphere are distorted by the ionosphere at centimetre and meter wavelengths, which makes low-frequency radio astronomy, especially at $\lesssim 1$\,GHz, particularly challenging. The three-dimensional structure of the ionosphere introduces phase errors into incoming cosmic signals recorded by ground-based radio telescopes. To obtain sensitive observations at these frequencies, it is critical to calibrate and remove ionospheric corruption from the measured data. \\
The ionosphere is a partially ionized plasma layer extending from about 50\,km to beyond 1000\,km altitude, primarily ionized by solar X-ray and extreme ultraviolet (EUV) radiation. While cosmic rays contribute to ionization, their contribution is significantly smaller than that of solar radiation. At nighttime, due to the recombination process, the number of electron-ion pairs decreases. Since there is no solar radiation to offset this loss through new ionization, the plasma density drops significantly, particularly in the lower regions. As a result, the electron density peak lies at an altitude of about 250 -- 500\,km \citep[see][]{Mannucci1998RaSc...33..565M}.\\
Radio interferometers are especially sensitive to ionosphere-induced path length differences. For effective calibration, the required antenna-based, time- and direction-dependent phase corrections typically must be determined with high precision using self-calibration or field-based techniques applied directly to interferometric observations. At frequencies around $200$\,MHz, this often demands ionospheric calibration at high temporal resolution, on the order of ten seconds. The configuration of interferometric arrays, along with the spread of calibrator sources across the sky, provides dense instantaneous sampling of the ionosphere’s spatially varying electron density. This sampling can be far finer than what is achievable even with dense grids of GPS receivers. As a result, radio interferometers offer a powerful synergy: they not only allow removal of ionospheric distortion from astronomical data but also contribute valuable information to ionospheric science at spatial scales smaller than those accessible through GNSS alone. \\
Radio interferometers are also significantly more sensitive to electron density changes than modern GNSS receivers. Even assuming a phase stability of about ten degrees for instruments such as the Square Kilometre Array (SKA), their sensitivity between 70 and 1420\,MHz corresponds to roughly 1 to 10 mTEC. This represents an improvement of about an order of magnitude over GPS-based experiments, and interferometric studies have even led to the discovery of new classes of travelling ionospheric disturbances (TIDs). \\
For single-dish radiometers, ionospheric effects appear through refraction, absorption, and emission, which can corrupt all-sky averaged cosmological signals at low frequencies. Ionospheric electron density fluctuations are strongly influenced by solar activity at various timescales, and solar phenomena, including radio bursts and sunspot variations, that exhibit flicker-noise ($\frac{1}{f}$) behaviour. These fluctuations impose additional corruption on single-antenna measurements, making high time resolution ionospheric calibration essential and placing stringent requirements on signal-to-noise ratios for successful calibration \citep[see][and references therein]{Datta2017}.\\
Over the past decade, ionospheric characterization and mitigation techniques have been developed using current radio interferometers. The Very Large Array\,(VLA) pioneered direction-dependent calibration at 74 MHz, while the Low-Frequency Array\,(LOFAR) demonstrated sub-kilometre-scale ionospheric imaging with TEC measurements approximately ten times more accurate than GNSS \citep{Intema2009A&A...501.1185I, Helmboldt2012RaSc...47.5008H, Helm2012RaSc...47.0L02H_spectral}. The Murchison Widefield Array (MWA) Phase II characterized ionospheric turbulence with 1–10\,mTEC sensitivity, revealing 
$\sim 50\%$ of ionospheric screens exhibit significant non-linear structures at sub-kilometre scales \citep{Loi2015RaSc...50..574L,Rioja2018, Rioja2022}. The upgraded Giant Metrewave Radio Telescope (uGMRT) enabled multi-frequency observations \citep{mangla2023exploring, Banerjee2025} to characterize both ionospheric structures and TIDs. However, current instruments face fundamental limitations in collecting area, baseline distribution, and temporal resolution.\\
The upcoming SKA \citep[see][]{Dewdney2009IEEEP..97.1482D, Braun2019arXiv191212699B}, now under construction at two strategic sites: Australia and South Africa, represents a transformative leap in ionospheric observation and calibration. The lower-frequency part, SKA-Low, located in Western Australia, consists of phased dipole arrays operating between 50 and 350\,MHz. This array will initially comprise $\text{AA}^*$ (307 stations), expanding to the full $\text{AA}4$ configuration of 512 stations. Its design features a dense $\sim 1-2$ km core containing $50\%$ of its stations. The remaining stations are distributed logarithmically along three spiral arms, extending to a maximum baseline of 74\,km. On the other side, SKA-Mid, located in South Africa, utilizes parabolic dishes optimized for higher-frequency observations, spanning $350\,\text{MHz}$ to $14\,\text{GHz}$ across multiple bands. The initial $\text{AA}^*$ configuration consists of 144 dishes (integrating 64 MeerKAT dishes), growing to the $\text{AA}4$ design of 197 dishes, extending to a maximum baseline of $150\,\text{km}$. The SKA requires precise calibration techniques to calibrate both instrumental and atmospheric propagation contributions as functions of time, frequency, and position, to meet its ambitious scientific goals. The development of advanced calibration algorithms is an active field of research, with the efforts focused on characterising the spatial and temporal scales of ionospheric structures. This complementary dual-array architecture is revolutionary for ionospheric studies. The SKA-Low telescope, with its heightened sensitivity at lower frequencies where ionospheric effects are most significant, allows for the detection and detailed analysis of ionospheric structures with exceptional accuracy. Meanwhile, SKA-Mid operates at higher frequencies with enhanced angular resolution, facilitating the validation of ionospheric models and the characterization of disturbances at multiple scales. Together, these instruments offer comprehensive, multi-frequency data collection of electron density over extensive baselines, providing an unparalleled ability to study ionospheric physics.

Next, we review the current state of ionospheric studies using first-generation telescopes. In Section~\ref{sec:radio_interferometer_ionosphere}, we present the ionospheric effects on electromagnetic radiation at low radio frequencies and their impact on radio-interferometric data. We then describe antenna-based (Section~\ref{sec:antenna_based_method}) and field-based (Section~\ref{sec:field_based_method}) reconstruction methods, followed by scintillation effects (Section~\ref{sec:scintillations}). In Section~\ref{sec:simulation} we discuss simulations of ionospheric-corrupted SKA data, and in Section~\ref{sec:future_SKA} we forecast calibration challenges for SKA-Low and SKA-Mid. We conclude in Section~\ref{sec:summary}.

\section{Effect of the ionosphere on radio interferometers}
\label{sec:radio_interferometer_ionosphere}
A radio interferometer measures the spatial coherence function by cross-correlating signals from pairs of antennas observing the sky \citep[see][]{Cornwell1999ASPC..180..187C}. This cross-correlation, known as the visibility, encodes information about both the far-field radiation pattern and the sky’s intensity distribution. The sky brightness distribution (i.e., the radio image) can be reconstructed by performing a Fourier inversion of the measured visibilities. Specifically, the observed visibility ($V_{ij}$) for an antenna pair (baseline $i,j$) is defined by the van Cittert-Zernike theorem:
\begin{equation}
V_{ij} = G_i \left( \iint_{lm} \frac{1}{n} \vec{E}_i \mathbf{B} \vec{E}_j^{H} e^{-2\pi i (u_{ij} l + v_{ij} m + w_{ij}(n-1))} \, dl \, dm \right) G_j^{H}
\label{eq:visibility}    
\end{equation}
Here, $B$ is the intrinsic source brightness matrix, $G_i$ represents the direction-independent effects\,(DIEs) or the $uv$-Jones terms, and $E_i$ is the direction-dependent effects\,(DDEs) or sky-Jones terms. \par
\begin{figure*}[ht]
    \centering
    \includegraphics[width=.9\columnwidth]{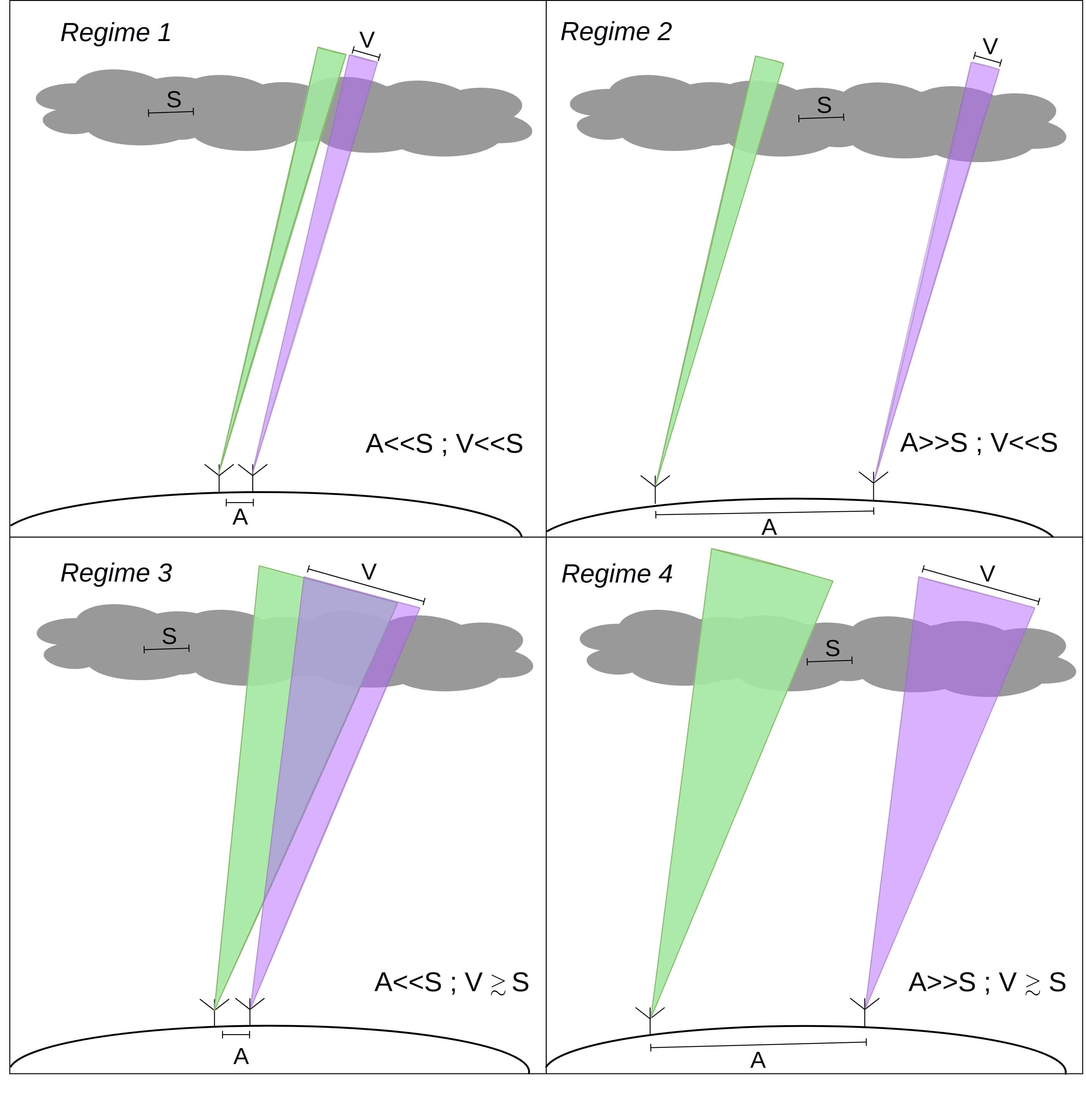}
    \caption{Schematic overview of four calibration regimes discussed by \citet{Lonsdale2005ASPC..345..399L} for low-frequency arrays. The quantities A, S, and V are the array size on the ground, the ionospheric irregularities scale size, and the FoV at projected ionospheric altitude, respectively. For isoplanatic conditions (V$<<$S; Regime 1 and 2), the ionospheric phase rotation does not vary much within the FoV of each antenna. For non-isoplanatic conditions (V$\gtrsim$S; Regime 3 and 4), the ionospheric phase rotation per antenna varies over the FoV. Adapted from \citet{mangla2023exploring}. }
    \label{fig:four_regimes}
\end{figure*}
The highly dynamic nature of the ionosphere presents a significant challenge in determining its effects on low-frequency arrays. Calibrating for this continuously varying medium is considerably more complex than correcting for slowly changing instrumental gain. Effective ionospheric calibration requires that the parameters describing the ionosphere above the array be sufficiently constrained so that the interferometric phase can be accurately measured and corrected. This correction must account for the phase as a delay on all baselines, across the entire field of view (FoV), and throughout the observation duration. To address this, \citet{Lonsdale2005ASPC..345..399L} established a framework based on the relationships between three critical length scales: (a) the physical scale size of ionospheric fluctuations\,(S) that induce noticeable phase delays, (b) the physical scale of the array\,(A) on the ground, and (c) the projected size of the FoV\,(V) of a single antenna at a typical ionospheric altitude. This yields four calibration regimes (see Fig. \ref{fig:four_regimes}):
\begin{itemize}
    \item In Regimes 1 (A $<<$ S; V $<<$ S) and 2 (A $>>$ S; V $<<$ S), the relatively narrow beam patterns ensure that the TEC is approximately `constant' across FoV. Since phase rotation is direction-independent, ionospheric effects can be effectively mitigated via self-calibration with sufficient temporal resolution.
    
    \item Regime 3 (A $<<$ S; V $\gtrsim$ S), wide FoVs induce significant TEC variations across the FoV. In compact arrays, TEC variation across antennas forms an approximate linear gradient, causing time- and direction-dependent source position shifts without shape deformation. Calibration here is simplified, requiring few parameters because all antennas share the same FoV region.

    \item Regime 4 (A $>>$ S; V $\gtrsim$ S), wide FoVs also induce significant TEC variations across across the FoV. For extended arrays, TEC variation becomes highly complex and non-linear, deviating substantially from a simple gradient. The lines of sight\,(LoS) of different antennas to the same source diverge significantly, resulting in both source position shifts and deformations. This scenario requires a robust, direction-dependent calibration framework capable of modeling a full three-dimensional ionospheric phase structure. 
\end{itemize}

The primary systematic effect introduced by the ionosphere in radio interferometry is the propagation delay, which stems from the varying refractive index ($n$) of the ionospheric plasma along the wave's path. The total propagation delay integrated along the LoS at a given frequency $\nu$ is expressed as a phase rotation:
\begin{equation}
    \label{eq:phase_rotation}
    \phi_{\rm ion} = - \frac{2 \pi \nu}{c} \int (n - 1)\,dx
\end{equation}
Assuming a refractive index\,($n$) that is constant throughout space and time, the ionosphere would introduce only a constant phase offset. This would manifest as a uniform spatial shift of the observed sources relative to the true sky.\par
By assuming a cold, collisionless, magnetized plasma, which is the typical characteristics of the Earth's ionosphere, one can calculate an exact expression for the refractive index $n$ \citep{davies1990ionospheric_book}. In the high-frequency limit, where the observing frequency $\nu$ is significantly greater than the plasma frequency $\nu_p$ (which is approximately $\rm\sim\,10\,MHz$), this expression can be simplified. Following the approach of \citet{Datta2008RaSc...43.5010D_higherorderionosphericerror}, a third-order Taylor expansion allows the refractive index to be approximated by retaining terms up to $\nu^{-4}$:
\begin{equation}
    n \approx 1 - \frac{q^2}{8\pi^ 2m_e \epsilon_0} \cdot \frac{n_e}{\nu^2} \pm \frac{q^3}{16 \pi^3 m_e^2 \epsilon_0} \cdot \frac{n_e B \cos\theta}{\nu^3} \\- \frac{q^4}{128 \pi^4 m_e^2 \epsilon_0^2} 16\cdot \frac{n_e^2}{\nu^4} - \frac{q^4}{64 \pi^4 m_e^3 \epsilon_0} \cdot \frac{n_e B^2 (1+\cos^2\theta)}{\nu^4}
    \label{eq:refractive_index_expand}
\end{equation}
where $q$ is the electron charge, $m_e$ is the electron mass, $\epsilon_0$ is the electric permittivity in vacuum, $n_e$ is the number density of free electrons, the magnetic field strength is denoted by $B$, and $\theta$ is the angle between $B$ and the propagation direction of an electromagnetic wave. The first term is the dominant factor, responsible for the dispersive delay proportional to the TEC along the LoS. For signal frequencies exceeding a few hundred $\text{MHz}$, the subsequent higher-order terms can typically be neglected. The second term describes the effect of Faraday rotation, where the sign convention distinguishes between left-hand and right-hand polarized signals. While the last two terms are usually ignored in general radio astronomy, they become essential for accurately modeling the ionosphere at very low observing frequencies, specifically below $40\,\text{MHz}$ \citep[see][]{Hoque2008RaSc...43.5008H}.
Uncorrected ionospheric phase errors severely impact image quality. This causes an apparent shift in source positions, and when higher-order terms are significant, it leads to distortion or even the suppression of sources. Precise ionospheric phase calibration is therefore essential for accurately localising and characterising astrophysical sources. Moreover, the measured phase provides a powerful diagnostic for studying the ionosphere's dynamics, provided calibration is performed with adequate temporal, spatial, and directional precision—most effectively achieved through self-calibration or field-based techniques. Modern radio interferometers, thanks to their array geometry and calibrator distribution, offer a much finer and more densely sampled map of the spatially variable ionospheric electron density than GNSS receiver grids, enabling simultaneous advances in interferometric imaging and ionospheric research \citep[see][]{Kassim2010amos.confE..59K}. A key advantage of radio interferometers over GNSS receivers is their superior sensitivity to subtle ionospheric electron density fluctuations. These fluctuations are primarily driven by solar radiation, with electron density variations linked to diverse solar disturbances that exhibit flicker-noise\,(1/f) characteristics over different timescales \citep{Datta2016}. This dynamic ionospheric corruption in raw antenna signals necessitates ionospheric calibration at high temporal resolution. Meeting this demand requires a high signal-to-noise ratio (SNR) for successful and accurate phase correction, essential to mitigate atmospheric effects in low-frequency radio astronomy.

\section{Exploring ionospheric structures using Antenna-based Method}
\label{sec:antenna_based_method}
In the antenna-based method, ionosphere measurements are derived in such a way that a single bright source is observed at the phase centre, and the time-dependent phase variation is measured on every baseline (the separation between a pair of antennas). The electromagnetic signals from radio sources are perturbed as they propagate through the ionosphere. The dominant, first-order phase term ($\Phi_{1st}$), approximated using Equations \ref{eq:phase_rotation} and \ref{eq:refractive_index_expand}, is given by:

\begin{eqnarray}
    \label{eq:phi2tec}
    \phi_{\rm ion} = 84.36 \left ( \frac{\nu}{\mbox{\scriptsize 100 MHz}} \right )^{-1} \left ( \frac{\mbox{\scriptsize TEC}}{\mbox{\scriptsize 1 TECU}} \right ) \mbox{ radians}
\end{eqnarray}
Equation \ref{eq:phi2tec} establishes a direct proportionality: the measured ionospheric phase is proportional to the difference in TEC ($\Delta \text{TEC}$) along the LoS between the two antennas. This technique is formally designated as the antenna-based method, and it is specifically designed to investigate the temporal variation in differential TEC corresponding to the array's footprint in the ionosphere. This methodology has been successfully employed in numerous studies dedicated to ionospheric probing.
\begin{figure*}
    \centering
    \includegraphics[width=0.9\columnwidth]{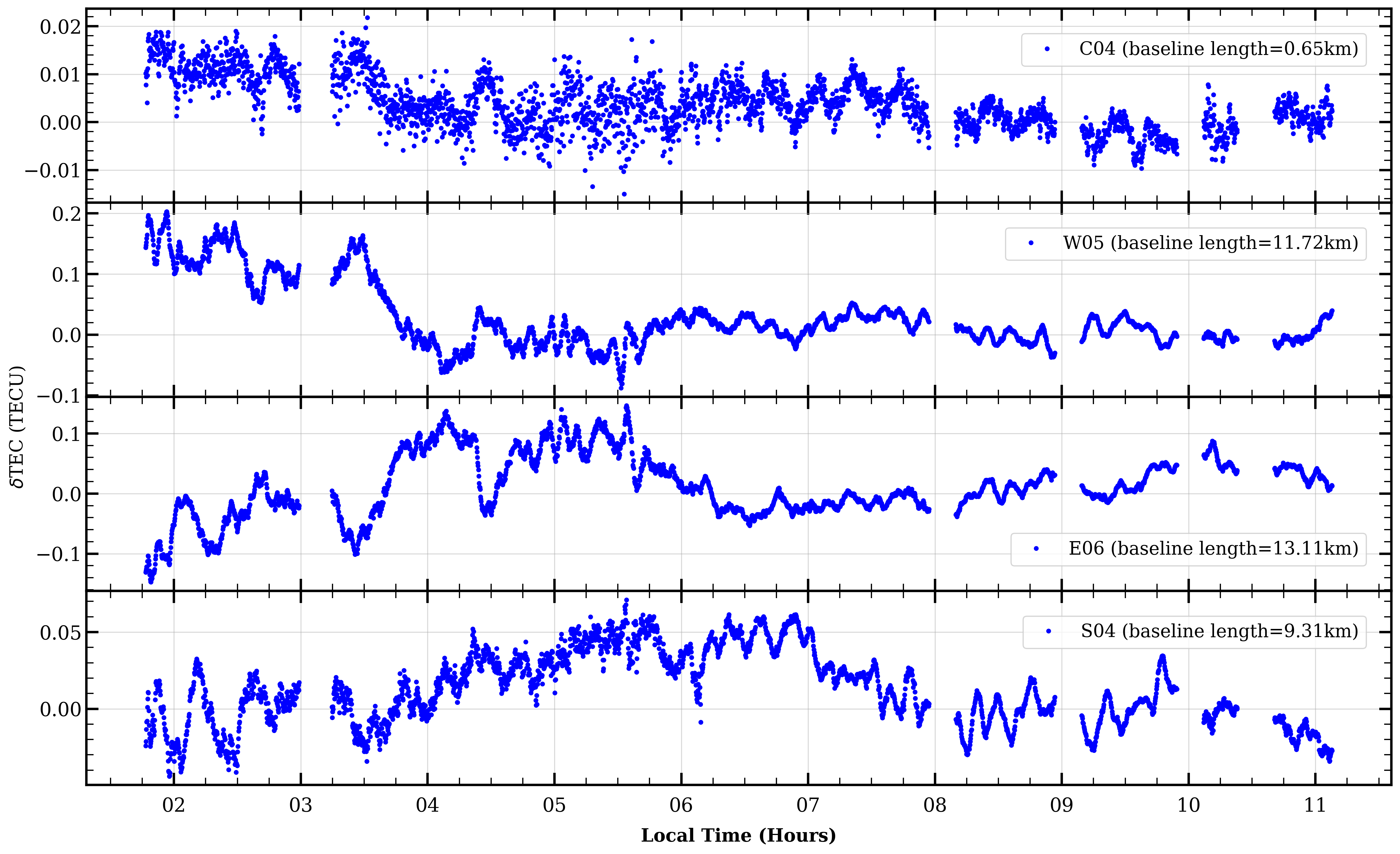}
    \caption{The differential TEC ($\delta \rm TEC$) variations are plotted as a function of time for multiple antennas located in the Central Square and along the arms of the GMRT. The $\delta \rm TEC$ values were computed along each antenna’s line of sight relative to the line of sight of the designated reference antenna \citep[see][]{mangla2023exploring}}
    \label{fig:centralsquare}
\end{figure*}
The high sensitivity of the VLA was leveraged by \citet{helmboldt2012VLAtec} through the observation of a single bright source (Cygnus A). This work demonstrated the ability to detect TEC variations with amplitudes below $10^{-3}\,\text{TECU}$ and to measure TEC gradients with an accuracy of approximately $0.2\,\text{mTECU}\,\text{km}^{-1}$. Following this, the complementary work by \citet{Helm2012RaSc...47.0L02H_spectral} focused on the spectral analysis of these TEC gradient measurements and detected and classified MSTIDs.

%
LOFAR observations of the bright quasar 3C 196 enabled \citet{mev16} to probe ionospheric structure with high precision ($10^{-3}\,\text{TECU}$). Their analysis of long-range baseline data, collected during winter nights (2012–2013), revealed anisotropic TEC irregularities and a power-law spectrum consistent with Kolmogorov turbulence in the ionosphere. Building on this, \citet{gasp18} quantified the impact of scintillation, showing that it significantly corrupts visibility amplitudes at ultra-low frequencies, resulting in an average of $30\%$ data loss during the night relative to the day. Consequently, this study emphasized the need for daytime observations, particularly for projects like the LOFAR Epoch of Reionisation (EoR) experiment. Further research involving a combined analysis of data from LOFAR, GNSS, and ionosondes \citep[see][]{Rich2020JSWSC..10...10F} uncovered a mechanism for structure evolution: when large-scale and small-scale TIDs travel orthogonally, they generate instabilities that facilitate the breakdown of large-scale structures into smaller components.
As one of the antenna-based observations, Figure~\ref{fig:centralsquare} represents GMRT observations of the astronomical source 3C\,68.2 at $235\,\text{MHz}$, recorded on August 6, 2012, covering the time from local midnight to post-sunrise. The dataset was pre-processed, flagged, and calibrated using the Common Astronomy Software Applications (\texttt{CASA} \footnote{\texttt{CASA} is a python-based software application package for Radio Interferometer data processing \cite{The_CASA_Team_2022}; \url{https://casa.nrao.edu/}}), following the detailed procedure presented in \citet{mangla2022mnras}. 
The complex antenna gain solutions contain both the ionospheric phase term and instrumental noise. To isolate the ionospheric component, we employed the continuum subtraction method \citep[see][]{mangla2022mnras,helmboldt2012VLAtec}. The resulting ionospheric phase was then converted to differential Total Electron Content ($\Delta \text{TEC}$) using Equation \ref{eq:phi2tec}, with the time series for four antennas baselines shown in Figure~\ref{fig:centralsquare}. To quantify the precision of our measurement, the uncertainty was estimated using the Median Absolute Deviation (MAD) of the $\Delta \text{TEC}$ time series at each step. This MAD computation included four adjacent time steps to ensure higher accuracy. The resulting uncertainty in $\Delta \text{TEC}$ is approximately $1 \times 10^{-3}\,\text{TECU}$, and these MAD values are also plotted to illustrate the relative accuracy achieved. Analysis of the arm antennas reveals a critical dependence: the $\Delta \text{TEC}$ along each arm, when normalised by the baseline length (with respect to reference antenna 'C06'), is proportional to the baseline length. Crucially, the observation of distinct patterns along the three arms confirms that the ionosphere varies significantly in both space and time. This spatial and temporal variability suggests the propagation of multiple waves or a dominant wave accompanied by smaller, differently directed components. 

Spectral analysis of these patterns, following the methodologies of \citet{Helm2012RaSc...47.0L02H_spectral} allows for the extraction of the speed, direction, and physical size of the dominant ionospheric structure(s) across the full array.
%
Due to GMRT's sensitivity to the differential TEC ($\Delta \text{TEC}$) between antenna pairs, measuring the resulting TEC gradient is essential for a complete understanding of ionospheric fluctuations. To transition from the measured slant $\Delta \text{TEC}$ to the desired vertical $\Delta \text{TEC}$, a geometric correction is first performed based on the thin-shell approximation of the ionosphere, located at a single 'peak height'. This peak height is determined using the International Reference Ionosphere\,(IRI) Plas model \footnote{IRI extended to Plasmasphere \url{http://www.ionolab.org/iriplasonline/}} \citep{Gulyaeva2012},
which leverages the GMRT's location and observation time. Notably, the IRI Plas model is preferred because, over the Indian longitude sector, it provides the most accurate estimation of electron density and TEC under disturbed conditions compared to other models, owing to its inclusion of electron density distribution up to the plasmasphere \citep[see][and references therein]{sc38}. After this well-documented correction \citep[see][Appendix A]{mangla2022mnras} is applied to the $\Delta \text{TEC}$ time series for every antenna, the TEC gradient across the full GMRT array is calculated.
\begin{figure}
    \centering
    \includegraphics[width=0.9\columnwidth]{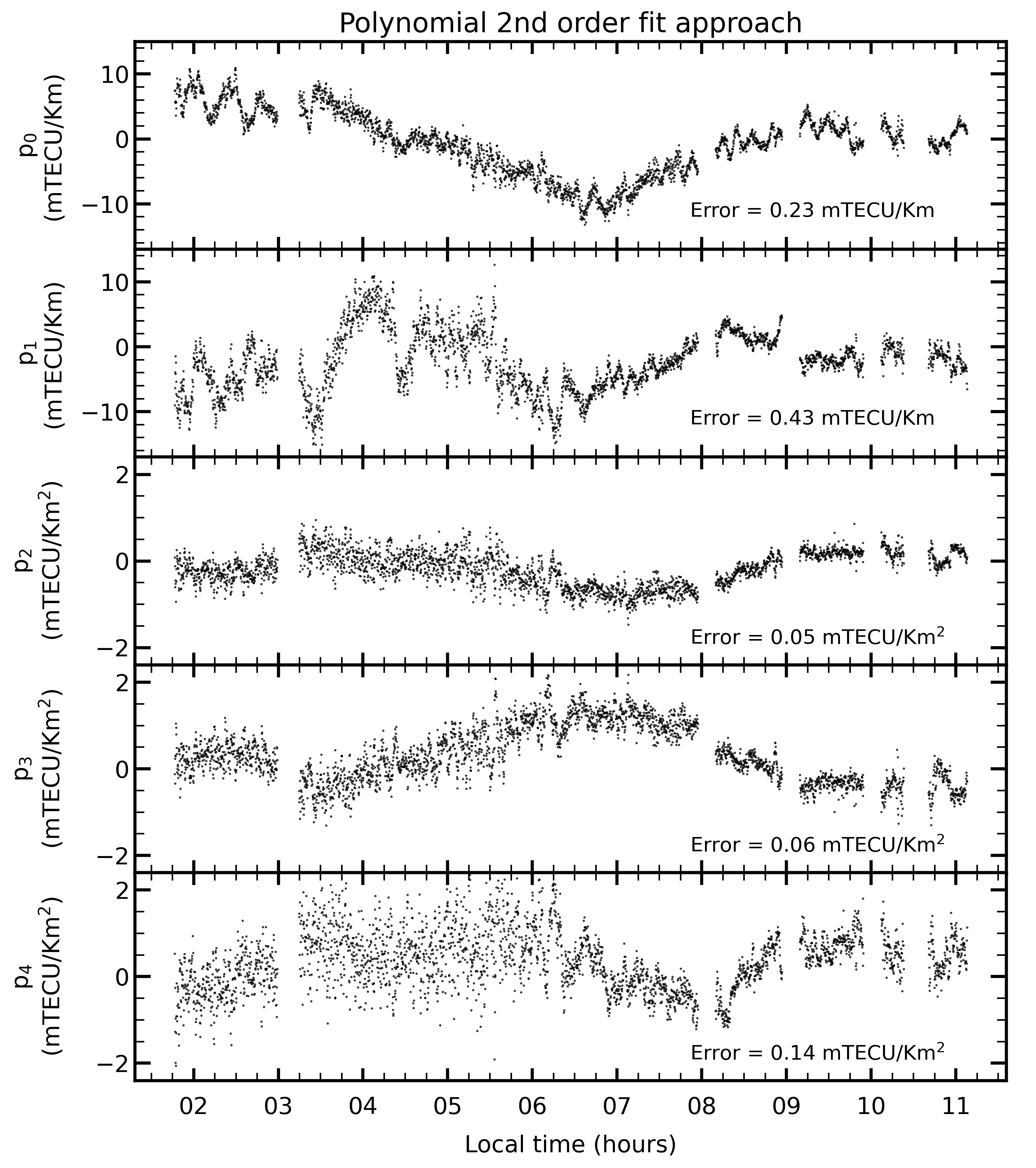}
    \caption{The fitted coefficients ($\rm p_0$ to $\rm p_4$) along the observation. These coefficients are fitted using the second-order polynomial equation independently for each time step using each of the antenna pairs. The estimated error for each coefficient is also mentioned in the respective panel. Adapted from \citet{mangla2023exploring}}
    \label{fig:polynomial}
\end{figure}
The two-dimensional TEC surface over the array can be computed using the second-order two-dimensional Taylor series (maximum baseline for the array is smaller than transient ionospheric waves), which has the following form:
\begin{equation}
    \label{eq:poly_2nd}
    \mbox{TEC} = p_{0}\,x + p_{1}\,y + p_{2}\,x^{2} + p_{3}\,y^{2} + p_{4}\,x\,y + p_{5}
\end{equation}
where x and y are antenna positions along with north-south and east-west directions. $\rm p_0$ to $\rm p_5$ are the polynomial coefficients. By calculating the difference of \dtec between antenna pairs (120 baselines) at each time step, one can increase the accuracy of these polynomial coefficients/parameters. Thus, equation \ref{eq:poly_2nd} transforms into the following form:
\begin{eqnarray}
    \label{eq:dpoly_2nd}
    \delta \mbox{TEC}_i - \delta \mbox{TEC}_j \! &=& \! p_0\,(x_i-x_j)+p_1\,(y_i-y_j)\nonumber \\ 
    &+& \! p_2\,(x_i^2-x_j^2) + p_3\,(y_i^2-y_j^2) \nonumber \\
    &+& \! p_4\,(x_i\,y_i-x_j\,y_j)
\end{eqnarray}
where subscripts $i$ and $j$ correspond to different antenna pairs (where $i\,\neq\,j$). It is important to note that the fitting is done independently at each time step, thus conserving the temporal and spatial TEC variation over the array. The obtained polynomial coefficients are shown in Fig.\ \ref{fig:polynomial} along with the standard error for each coefficient. One can easily notice that the amplitude for the $\rm p_1$ coefficient (along the east-west direction) is significantly high during the local nighttime. The same phenomenon is observed for $\rm p_0$ coefficient (along north-south direction) during the sunrise hour (around 6:00\,AM), which is a known behaviour of MSTIDs, commonly detected around sunrise and sunset time \citep[see][]{her06}. Variations in higher-order coefficients ($\rm p_2$ to $\rm p_4$) are more significant during the nighttime, suggesting unanticipated ionospheric changes in and around the EIA region during these local times. \par
\subsection{Extending Antenna-Based Measurements to Map Ionospheric Phase Screen}
In extension to the above, the Ionospheric Phase Screen ($Z$) was reconstructed by utilising the \texttt{grad2surf} algorithm by \citet{Harker2014}. That algorithm is primarily designed to reconstruct a 3D surface of an image with a gradient field that is measured with Photometric Stereo \citep{woodham1980photometric}. \texttt{grad2surf} algorithm requires the gradient fields $\hat{Z}_x$ and $\hat{Z}_y$, respectively measured in $x$ and $y$ direction. 
\begin{equation}
  \frac{\partial Z}{\partial x}=ZD_x^T\approx\hat{Z}_x
\end{equation}
\begin{equation}
  \frac{\partial Z}{\partial y}=D_yZ\approx\hat{Z}_y
\end{equation} 
Here, $Z$ is the actual matrix of unknown surface (phase screen) with unknowns $z_{ij}$, $D_x$ and $D_y$ are differential matrices and provide the theoretical/numerical gradient field of the surface, whereas terms after   $\approx$, $\hat{Z}_x$ and $\hat{Z}_y$ provide the measured gradient field. Thus, the matrix Frobenius norm can be used to express the global least squares cost function for reconstructing a surface from its gradient field as follows:
\begin{equation}
    \epsilon Z = \left\| Z D^T_x - {\hat{Z}}_x \right\| _F^2 + \left\| D_yZ - {\hat{Z}}_y \right\| _F^2,
\end{equation}
This is the Euclidean distance between the gradient field of an unknown surface $Z$ and the measured gradient field. From a mathematical perspective, this cost function is a function of the unknown function (surface) $Z$ and can be seen as a discrete function with respect to the calculus of variations. The effective normal equations of the least-squares problem are obtained by differentiating with respect to the matrix $Z$ to determine the minimum of the cost function.
\begin{equation}
    D^T_yD_yZ + ZD^T_xD_x - D^T_y{\hat{Z}}_y - {\hat{Z}}_xD_x = 0
\end{equation}
This matrix equation is a set of linear equations in the unknowns $z_{ij}$ and is known as a Sylvester Equation. This algorithm has been utilised to recreate a phase-screen map over GMRT by replacing $\hat{Z}_x$ and $\hat{Z}_y$ with polynomial-interpolated directional TEC gradients, respectively in $x$ and $y$ direction, and $Z$ represents the TEC map $TEC(x,y)$.
We use the methodology described in \citet{helmboldt2012VLAtec} to estimate the extra phase produced as a result of the radio signal travelling through the Earth's ionosphere. To obtain the ionospheric information, several procedures are carried out on the phase data once the phase terms from Common Astronomy Software Application (\texttt{CASA}) have been calculated. The ionospheric contribution, which is greatest in low-frequency regimes, is included in the phase terms along with additional effects.  \par
The phase difference between the two antenna elements \citep{Intema2009A&A...501.1185I,helmboldt2012VLAtec} is given by 
\begin{eqnarray}
        \Delta\phi = \Delta\phi_{\rm ion} + \Delta\phi_{\rm other}
    \label{eq:Total_phase_diff}
\end{eqnarray}
where $\Delta\phi_{\rm ion}$ represents the difference in the ionospheric phases of the two antennas along the line of sight given by equation (\ref{eq:phi2tec}), $\Delta\phi_{\rm other}$ denotes the combined differences in the instrumental effects between the two antennas, $2\pi$ ambiguities and the phase difference from the observed source structure. The  $\Delta\phi_{\rm other}$ phase delay effect is removed or reduced as the method mentioned in \citet{mangla2022mnras}.  From equation \ref{eq:phi2tec}, one can infer that the difference in TEC along the LoS to the two antennas is directly proportional to the measured ionospheric induced phase. This method will be referred to as the \textit{antenna-based method}, which can study the temporal variation in differential TEC ($\delta$TEC) corresponding to the projected array onto the ionosphere. Several studies have used this method to probe the ionosphere.
The resulting $\delta\rm{TEC}$ values are plotted for different antennas in the array with respect to reference antenna `C06' in Fig.~\ref{fig:centralsquare}. \citet{mangla2022mnras} proves the remarkable ability of the GMRT to detect small fluctuations while measuring TEC. These results suggest that the variation in $\delta\rm{TEC}$ for the southern arm is more than the other two arms and the central square configuration. 

\begin{figure}[!htbp]
    \centering
    \includegraphics[width=0.45\linewidth]{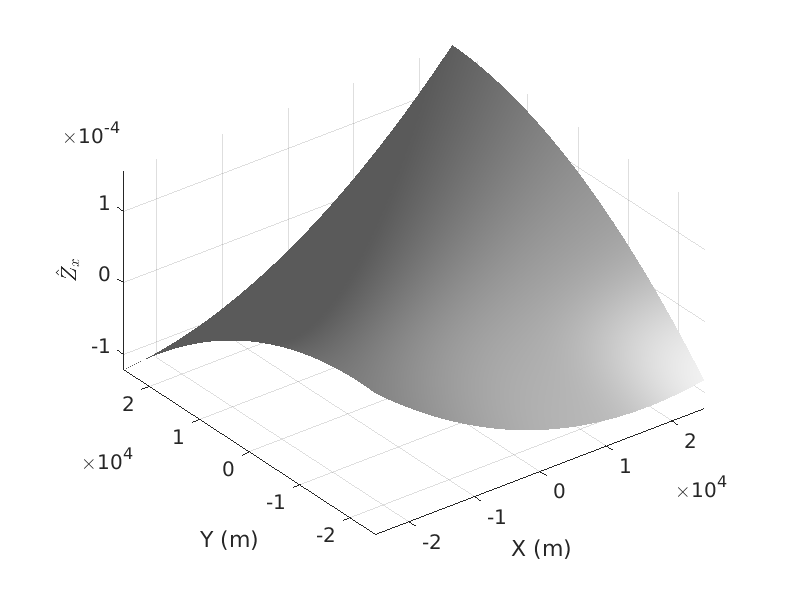}
    \includegraphics[width=0.45\linewidth]{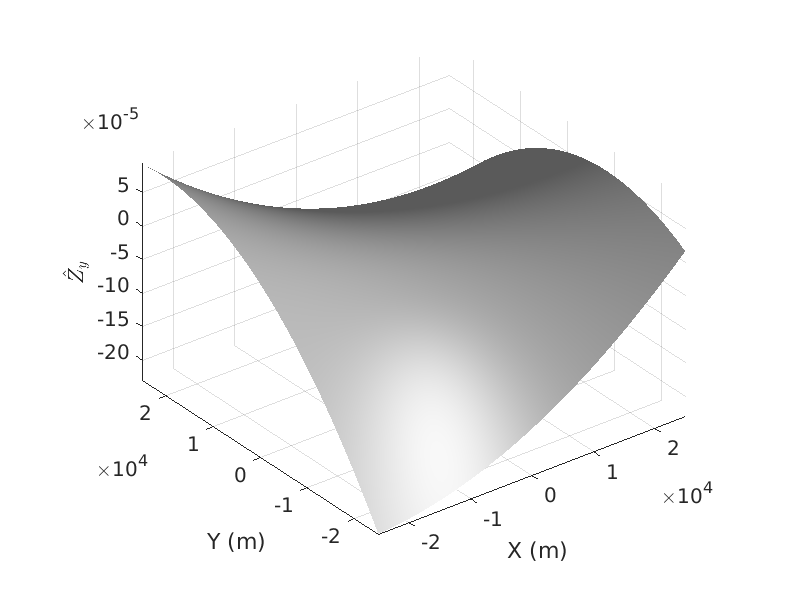}
    \caption{Figure shows measured TEC gradients $\hat{Z}_x$ and $\hat{Z}_y$ in $x$ and $y$ directions respectively \citep[Adapted from][]{Brawar2025}.}
    \label{fig:grad_map}
\end{figure}

Taking the GMRT antenna `C06' as the reference antenna, similar TEC gradients with all the active antennae were calculated along with time. The gradient field is then mapped over the GMRT area of $50\text{km} \times 50 \text{km}$ using a 3rd order polynomial fit onto a $100\times100$ pixel \citep[for more see][]{mangla2022mnras}. Then $\hat{Z}_x$ and $\hat{Z}_y$ in Figure \ref{fig:grad_map} represent the gradient measured in x and y direction with an interpolation polynomial fit.

Using these TEC gradients $\hat{Z}_x$ and $\hat{Z}_y$ with \texttt{grad2surf} provides a TEC map with a mean value of zero, so to get a near true TEC measurement, we need to provide boundary conditions or an integral constant. \texttt{grad2surf} relies on the mathematical principle of integration ($\int f(x)\,dx$), and obtaining the correct result requires specifying the integration constant ($C$). For this purpose, we used IRI-2016 products for the GMRT location as input for boundary conditions. The following figure \ref{fig:iri_grad} provides a comparison of the IRI-2016 generated TEC map with respect to the TEC map reconstructed from gradients.

\begin{figure}[!htbp]
    \centering
    \includegraphics[width=0.45\linewidth]{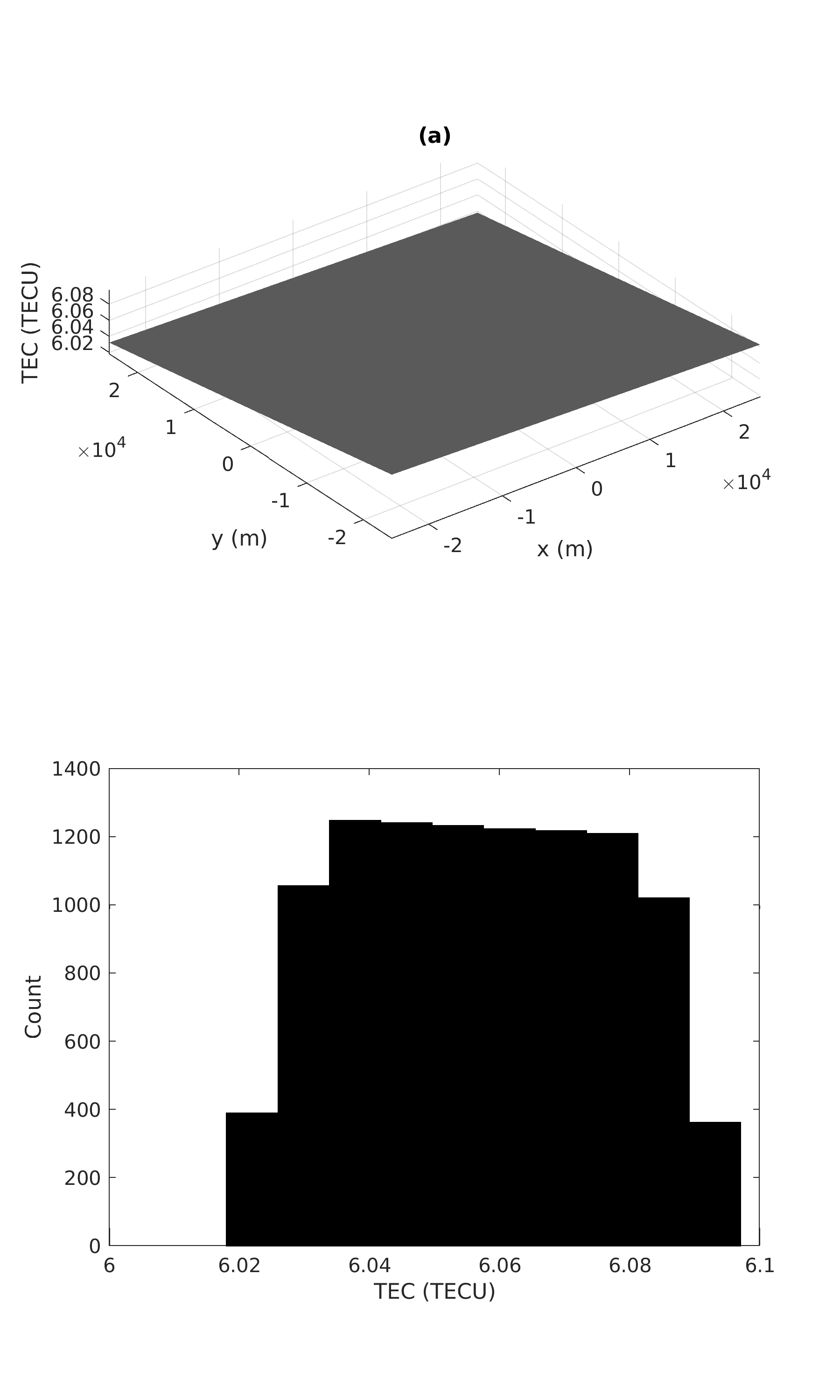}
    \includegraphics[width=0.45\linewidth]{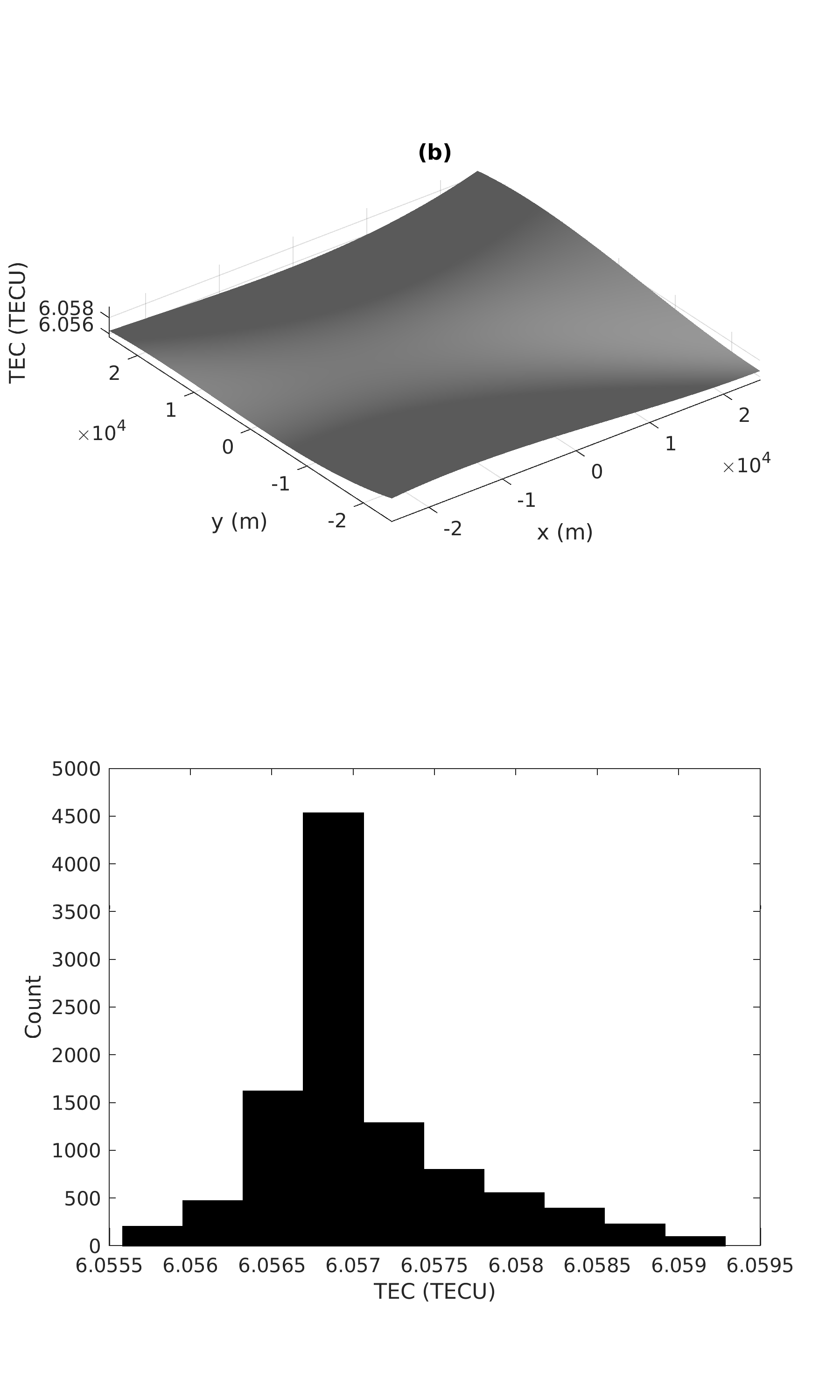}
    \caption{TEC map profile over the GMRT from (a)IRI-2016 and (b)Reconstruction from gradients during the observation period. Histograms provide the TEC fluctuation values within the area \citep[Adapted from][]{Brawar2025}.}
    \label{fig:iri_grad}
\end{figure}

\section{Exploring ionospheric structures using Field Based Method}\label{sec:field_based_method}
Unlike the antenna-based approach, wide-field radio interferometers are capable of simultaneously observing many celestial sources, resulting in a large collection of isotropically distributed ionospheric pierce points \citep[see][]{Cohen_2009, jordan2017iono_grad}. This mandates the use of the field-based calibration\,(FBC), where the phase is determined by quantifying the shifts in apparent source position instead of antenna-based solutions. The core principle of FBC is that the observed position shift of a source is directly proportional to the components of the TEC gradient along the LoS, averaged over the array and the snapshot time. Since the instrument measures differential phase between antenna pairs, the phase delay between the signal paths for a given pair translates into an apparent positional shift ($\Delta \theta$) in the source's actual location, defined by the angle:
\begin{equation}
    \label{eq:angle}
    \theta = \frac{\lambda}{L} \frac{\Delta\phi}{2\pi}
\end{equation}
where $\lambda$ is the observational wavelength, $L$ is the baseline length, and $\Delta\phi$ is the difference in ionospheric phase delays between pairs of antennas. If the ionospheric phase delay changes linearly across an array, all baselines ``observe'' the same position shift for a given source, and the image of a celestial source is merely moved from its true sky position. Using equations \ref{eq:phi2tec} and \ref{eq:angle}, the TEC gradient along a direction $r$ on the ground to the observed angular shift in the same direction $\theta_r$, is given by \citep[see][]{Cohen_2009,Helm2020RaSc...5507106H}:
\begin{equation}
    \label{eq:position_shift}
    \frac{d}{dr} {\rm TEC} = 1.2 \times 10^{-4} \left (\frac{\nu}{\rm100 MHz} \right )^2 \left ( \frac{\theta_r}{1''} \right ) \rm TECU\,km^{-1}
\end{equation}
If the ionospheric phase delay deviates from linearity across the array, and higher-order terms such as curvature become significant, the effect is two-fold: the apparent sky position will be shifted, and the image itself will become distorted.\par
The field-based method has been extensively utilized to probe ionospheric dynamics. For instance, \citet{Helmboldt2012RaSc...47.5008H} conducted a spectral analysis on the apparent positional shifts of 29 bright radio sources within a single VLA FoV at $74\,\text{MHz}$. This technique successfully detected wavelike ionospheric disturbances. Building on this approach, \citet{Loi2015RaSc...50..574L} investigated the ionosphere by studying the power spectrum of ionospheric TEC variations using the MWA, which were derived directly from radio source positional offsets. Their analysis clearly revealed the characteristic signatures of TIDs. \par
The technique of imaging ionospheric structure was utilized by \citet{Helm2020RaSc...5507106H}; they analysed 200 hours of data from the MWA's GLEAM survey \citep[see][]{Wayth2015PASA...32...25W}. Through spectral analysis of these resultant ionospheric images, the study uncovered key characteristics of nighttime activity, establishing a clear link: the generation of MSTIDs is associated with the formation of Sporadic E-layers.\par
The MWA employs advanced calibration techniques to mitigate ionospheric distortions, which are crucial for achieving its scientific goals, such as detecting the redshifted cosmological 21-cm signal from neutral hydrogen during the EoR \citep{2022Chege, 2021Chege}. The Real-Time System (RTS) calibration, for instance, provides high spatial and temporal resolution information about the ionosphere, allowing for the classification of ionospheric activity into four distinct types and the development of metrics for data quality assessment \citep{jordan2017iono_grad}. Recent studies with MWA have reported unusually strong ionospheric activity, including duct-like structures (roughly $50 \text{km}\times > 100$\,km), during magnetic storms, demonstrating its capability to capture ionospheric structures smaller than 100\,km \citep{2025Yoshiura}.\\
\section{Estimation of Amplitude Scintillation}
\label{sec:scintillations}
While the previous sections detailed the impact of phase errors, this section addresses ionospheric effects on the amplitude of radio interferometric visibilities.
When radio waves from compact sources pass through an ionized medium, they can be strongly affected by small-scale density structures in the electron density. This diffraction of the wavefront caused by small-scale irregularities leads to variations in the intensity of the wavefront as a function of distance from the scattering medium. This well-known phenomenon is scintillation. Combined with phase measurements discussed previously, scintillation observations reveal the spatial and temporal structure of ionospheric turbulence across scales from tens of meters to kilometres. The wide operational bandwidth of SKA precursor and pathfinder facilities, such as LOFAR, MWA, VLA, and uGMRT, is critical for accurately characterising the scattering regime (weak or strong) and developing methodologies for real-time ionospheric monitoring. The current and future radio interferometers are used to map the complex spatial, temporal, and frequency dependencies of ionospheric turbulence. This comparative analysis demonstrates how different array configurations and locations prioritise the study of distinct scale structures within the ionosphere.

\citet{Rich2020JSWSC..10...10F}, using LOFAR, observed ionospheric scintillation on the strong radio source Cassiopeia A at mid-latitudes across an observing bandwidth of 10–80\,MHz. Their work demonstrates LOFAR’s capability to characterise two simultaneous TIDs at different altitudes and propagating in different directions. Using delay-Doppler spectra (the 2-D FFT of the dynamic spectrum) analysis, they show two distinct parabolic arcs, indicative of scattering from two separate ionospheric layers: an F-region layer ($\sim 200-700$ km), consistent with a large-scale TID potentially driven by high-latitude geomagnetic substorms, moving northwest to southeast, and another a D-region layer ($\sim 60-70$ km), indicative of a smaller-scale disturbance likely caused by atmospheric gravity waves moving northeast to southwest. Complementary data from GNSS and ionosonde measurements confirmed the presence and behaviour of these TIDs. The spatial scales and velocities of plasma irregularities causing scintillation were estimated, highlighting the ability of LOFAR to probe small- to medium-scale ionospheric structures simultaneously at multiple altitudes. This simultaneous detection of two distinct scattering phenomena originating from different altitudes represents a crucial step in understanding ionospheric propagation. It demonstrates that the common assumption of a single, thin phase screen often fails to capture the true complexity of the ionosphere. The ionosphere must frequently be treated as a multi-layered or volumetric turbulent medium, where radio waves undergo complex ray tracing and refraction through superimposed disturbances.\\
Using MWA data, \citet{Waszewski2022PASA...39...36W} studied small-scale ionospheric structures by analysing weak amplitude scintillation measurements at 154\,MHz, focusing on mid-latitude ionosphere. Their work demonstrated MWA is sensitive to study small-scale ionospheric structures from Fresnel-scale structure on scales $\sim 300$\,meters up to a few kilometres. Their methods, adapted from interplanetary scintillation studies \citep[IPS;][]{Morgan2018}, effectively characterize fine plasma irregularities. The study found a strong correlation between scintillation indices and refractive ionospheric shifts, confirming that amplitude scintillation is a useful proxy for small-scale ionospheric turbulence. This approach extends the capability to probe ionospheric irregularities much smaller than those typically resolved by GNSS total electron content measurements, providing a detailed view of turbulent ionospheric conditions that impact low-frequency radio observations. Recent MWA extended baseline observations at the SKA-Low site revealed that 50\% of ionospheric screens show significant non-linear structures at scales exceeding 0.6\,km, with $1\%$ exhibiting sub-minute temporal variations; non-linear corrections are mandatory at 88 MHz during normal observations and at 154 MHz during poor conditions \citep{Rioja2022}.\\
The VLA's VHF system achieves high-precision measurements of ionospheric TEC gradients with a differential TEC precision of
$3\times 10^{-4}$\,TECU, spanning a much wider range of spatial scales (1–10\,km) than GNSS measurements. VLA observations provide TEC fluctuation spectra inaccessible to other techniques and have characterized ionospheric structure function properties for calibration algorithm development. \citet{helmboldt2012VLAtec, Helmboldt2012RaSc...47.5008H} detected and characterised medium-scale TIDs and identified quasi-periodic echo signatures linked to sporadic E-layer coupling mechanisms. VLA operates primarily at 74 MHz and has limited epochs; it lacks continuous survey capability and multi-latitude systematic characterisation.\\
The uGMRT occupies a unique geomagnetic location between the equatorial ionization anomaly (EIA) northern crest and the magnetic equator, making it exceptionally sensitive to ionospheric disturbances. At 235\,MHz, uGMRT measures TEC gradients with an order of magnitude higher sensitivity than GNSS, directly detecting ionospheric irregularities on scales from $\sim 100$\,meters to kilometers \citep{mangla2022mnras, mangla2023exploring}. Combined uGMRT, VLA, LOFAR, and MWA observations have been used to characterize ionospheric disturbances across multiple magnetic latitudes and provide forecasts for SKA performance.

In essence, SKA-Low AA* \& AA4 and SKA-Mid AA* \& AA4 will transform ionospheric scintillation studies from regional, limited-sensitivity snapshots into continuous, local, and high-fidelity observations, opening new frontiers in space weather research, plasma physics, and atmospheric-ionosphere coupling. This enables precision 21-cm cosmology by providing exquisite correction for ionospheric amplitude and phase corruption across unprecedented scales and frequencies.

\section{Simulation of Ionospheric effects}\label{sec:simulation}
To construct a physically motivated and realistic model of the ionosphere, several characteristic properties must be taken into account. The scale and structure of the electron content should be consistent with observations, and the electron distribution should evolve in a manner that is coherent in both space and time. Ionospheric irregularities often occur at the equator and high latitudes, whereas mid-latitude locations such as SKA-Low experience lower activity levels \citep{sc01}. Although the ionosphere is a dynamic, three-dimensional medium, it is often represented using a simplified two-dimensional thin-layer model \citep{Ratcliffe1956}. This approach approximates the ionosphere as a discrete spherical shell situated at a fixed altitude, $h_{\rm ion}$, above the Earth's surface. This approximation is motivated by the ionosphere's vertical structure, in which most free electrons reside in the F region, between 200\,km and 450\,km in altitude. Collapsing the full three-dimensional electron distribution onto a two-dimensional shell substantially simplifies modeling, while preserving essential behaviour such as large-scale spatial coherence and, to some extent, the elevation dependence of the projected electron content \citep[see]{martin2016}. In the thin-layer approximation, the ionosphere is described by a two-dimensional field of vertical total electron content\,(vTEC). For a given line of sight, the relevant TEC value is evaluated at the ionospheric pierce point\,(IPP), which is the location where the source direction vector intersects the TEC screen.

\begin{figure}
    \centering
    \includegraphics[width=\linewidth]{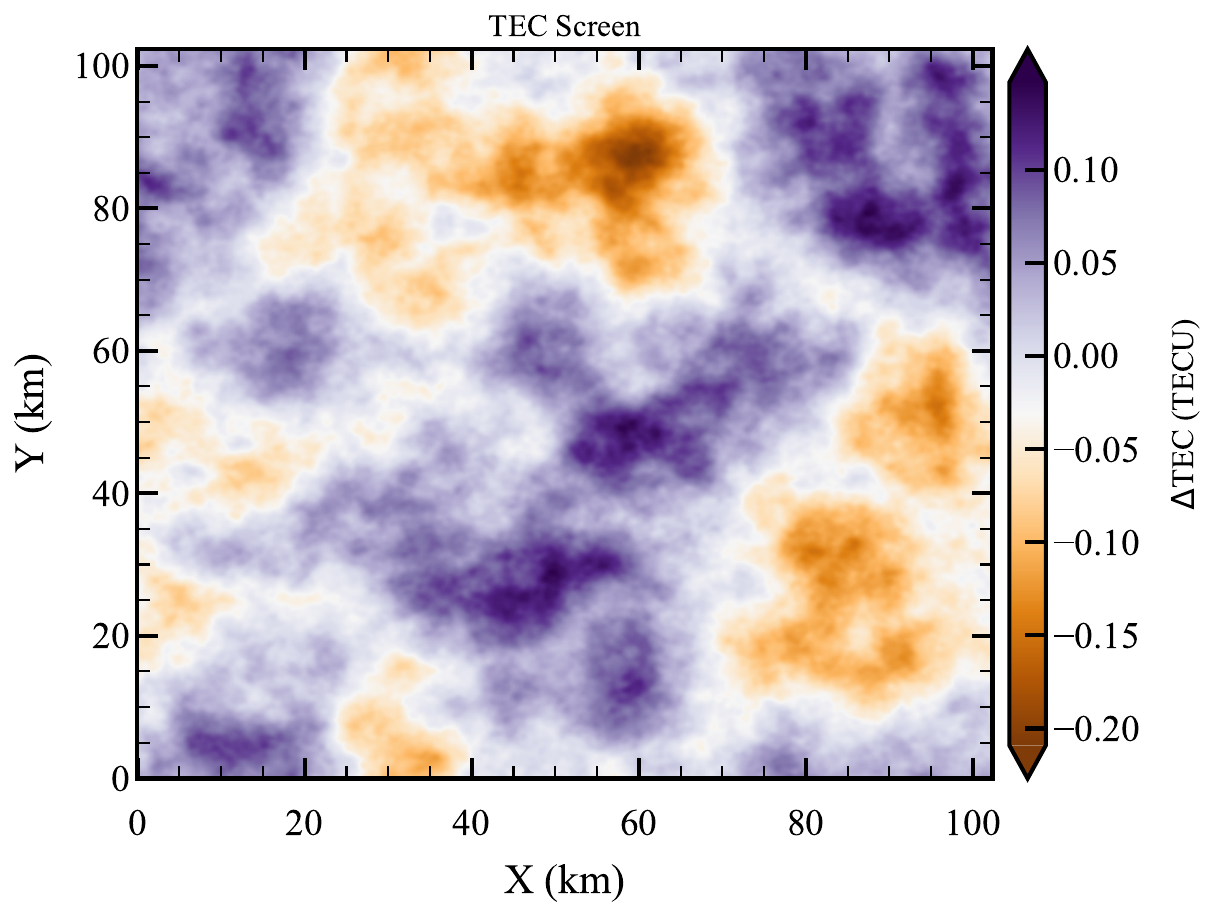}
    \caption{Simulated Kolmogorov's turbulence TEC screen was generated using the FFT algorithm. Adapted from \citet{Pal2025JCAP...02..058P}. }
    \label{fig:kolmo_screen}
\end{figure}

A substantial fraction of ionospheric inhomogeneity is attributed to turbulent processes \citep{materassi2019, giannattasio2018}. Such turbulence is commonly interpreted as a cascade in which energy is injected at large spatial scales and transferred progressively to smaller scales \citep{thompson2017}. Following this statistical framework, \citet{Pal2025JCAP...02..058P} implemented a time-evolving phase screen based on the Kolmogorov turbulence statistics. To simulate the temporal evolution of the medium, a "frozen" 2D phase screen is generated for a given epoch $t$, incorporating both the inner ($l_0$) and outer ($L_0$) scales of the turbulence. The power spectrum of the spatial phase fluctuations, $\Phi(\vec{k})$, is defined by:
\begin{equation}
    \left|\Phi (\Vec{k})\right|^2 \propto \left[k^2+ \left(\frac{1}{L_0}\right)^2\right]^{-\beta/2}\text{exp}\left(-\frac{k^2}{2/l_0^2} \right) ;   1/L_0 \ll  k \ll 1/l_0
\end{equation}
where $\Vec{k}$ is the spatial frequency. In the case of pure Kolmogorov turbulence, the spectral index $\beta=11/3$. At the LOFAR site, \citet{mev16} reported a slightly steeper spectrum of $\beta = 3.89 \pm 0.1$.
To anchor the model in observational reality, we estimate the characteristic scale structure using empirical ionospheric measurements from the SKA-Low site, based on MWA Phase-II extended baseline observations. Nevertheless, a more comprehensive characterization of the finite inner and outer scales at the Murchison Radio-astronomy Observatory\,(MRO) remains necessary. Observations from GLEAM indicate that the dominant ionospheric structures correspond to spatial scales of order hundreds of kilometres \citep{Helm2020RaSc...5507106H}. Figure~\ref{fig:kolmo_screen} presents an example of the simulated phase screen based on Kolmogorov statistics. 
 
There are several simulation tools available for modeling the physics of ionospheric turbulence, such as the \texttt{ARatmospy} package \citep{Srinath2015}. The \texttt{ARatmospy} (Auto-Regressive Atmosphere Generator) framework is capable of reproducing realistic ionospheric conditions by generating a time-evolving phase screen above the telescope site. This framework supports multi-layer phase-screen models at plausible elevations, each characterised by distinct propagation speeds and directions, while evolving smoothly in time to mimic the slowly varying nature of the ionosphere. This phase-screen model, integrated with the OSKAR simulator package, introduces direction-dependent calibration errors into the simulated visibilities. This OSKAR \citep{OSKAR2020} is the GPU–accelerated radio telescope simulator that was developed for its use in SKA since 2009. This simulation framework can be extended for tropospheric turbulence modeling to introduce direction-dependent calibration errors into the simulated visibilities relevant for the SKA-Mid. Such end-to-end simulations are essential for evaluating system performance, such as sensitivity and image fidelity, under realistic observing conditions. This also plays a key role in optimising survey strategies and identifying potential systematic effects in advance.

\section{Forecast for SKA}
\label{sec:future_SKA}
In this section, we have used the radio interferometric observations to estimate the ionospheric effects, which are dependent on the differential TEC values.  Based on the observations and existing literature, we forecast the phase errors that SKA-Mid and SKA-Low will be susceptible, as a function of observing frequencies. We have also translated them into the possible dynamic range limit in imaging.
At ultra-low frequencies, differential Faraday rotation (second term in equation\ \ref{eq:refractive_index_expand} and higher-order terms) will also become essential and cannot be ignored. Using LOFAR, \citet{gasp18} showed that higher-order ionospheric effects are only prominent for observations below $\sim$40\,MHz with a maximum baseline of $\sim$50\,km. As SKA-Low observational frequency starts from 50\,MHz, higher-order terms (third-order onwards) can be ignored for core antennas, but their effects cannot be ignored for longer baselines. \par
By substituting  equation \ref{eq:refractive_index_expand} into equation \ref{eq:phase_rotation}, one can estimate the effects caused by first and second order terms \citep[see][Chapter 9]{Petit2010ITN....36....1P} associated with dispersive delay and Faraday rotation, respectively:
\begin{align}
    \delta\Phi_1 &= -4840 \left( \frac{\nu}{100~\mathrm{MHz}} \right)^{-1}
    \left( \frac{\mathrm{dTEC}}{1~\mathrm{TECU}} \right)\; [\mathrm{deg}], \\[6pt]
    \delta\Phi_2 &= \pm 38 \left( \frac{\nu}{100~\mathrm{MHz}} \right)^{-2}
    \left[
        \left( \frac{\mathrm{dTEC}}{1~\mathrm{TECU}} \right)
        +
        \left( \frac{\mathrm{TEC}}{1~\mathrm{TECU}} \right)
        \left( \frac{\mathrm{d}B}{40~\mu\mathrm{T}} \right)
    \right]\; [\mathrm{deg}].
\end{align}
where we assumed magnetic field $B$\,=\,40\,$\mu$T with $\theta$\,=\,45$^\circ$. Under quiet-time ionospheric conditions, total TEC might change from $\sim$1\,TECU (during the nighttime) to $\sim$20\,TECU (during the daytime), affecting the second-order term. For example, consider a value for \dtec $\sim$0.3\,TECU, which is reasonable for a baseline of about $\sim$50\,km and observational frequency of 60\,MHz. The first-order term generates phase variations of several times of 2$\pi$. The second-order term, or the term due to Faraday rotation, produces an effect of around $\pm$ 50$^\circ$/75$^\circ$ assuming d$B$=1\% at night/daytime. This effect will not be negligible for SKA-Low and needs to be addressed carefully (see Table \ref{tab:ska_low} for more detailed information). SKA-Mid observational starts from 350\,MHz, second-order will be negligible compared to the dominating dispersive delay (first-order). As the maximum baseline length for SKA-Mid is 150\,km, \dtec values will be around 1.5\,TECU, which is considerably large and needs to be corrected (see Table \ref{tab:ska_mid} for detailed information). \par
%

\begin{table}[ht]
\caption{Typical ionospheric phase errors (in degrees) for SKA-Low}
\label{tab:ska_low}
\resizebox{\textwidth}{!}{%
\begin{tabular}{lcccccc}
\hline
 & \multicolumn{2}{c}{50 MHz} & \multicolumn{2}{c}{100 MHz} & \multicolumn{2}{c}{250 MHz} \\
\cline{2-7}
$\Delta$TEC (TECU) & I order & II order (day/night) & I order & II order (day/night) & I order & II order (day/night) \\
\hline \hline
1.0 (remote st., bad iono.)  & 9680 & 181/153 & 4840 & 45/38 & 1936 & 7/6 \\
0.4 (remote st., good iono.) & 3872 & 91/62  & 1936 & 23/15 & 774  & 4/2 \\
0.05 (across FOV)            & 484  & 38/9   & 242  & 9/2   & 97   & 2/$<$1 \\
0.02 (Core St.)              & 194  & 33/5   & 97   & 8/1   & 39   & 1/$<$1 \\
\hline
\end{tabular}%
}
\end{table}

\begin{figure}[t]
    \centering    
    \includegraphics[width=\linewidth]{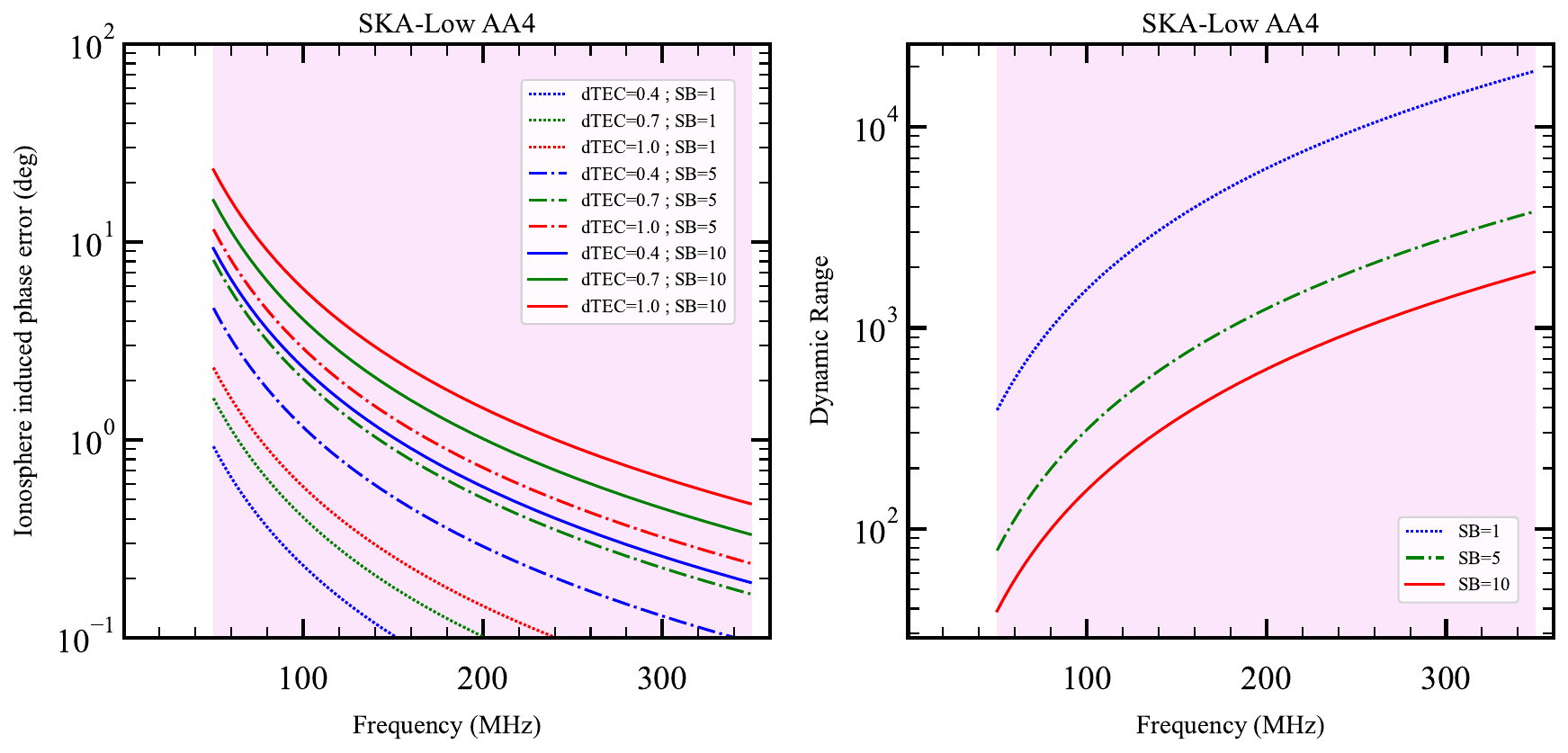}
    \caption{\texttt{Left:} Ionospheric-induced phase variations between the beginning and end of a band of 1, 5, and 10 SKA-Low AA4 sub bands ($1 {\rm SB} \simeq 0.006$\,MHz). Assumed \dtec is 0.4, 0.7, 1.0\,TECU. These values were considered for baseline lengths of a few tens of kilometers. SKA-Low frequency bandwidth is 50-350\,MHz. \texttt{Right: } Dynamic range ($DR$) as a function of frequency for sub bands 1, 5, and 10 ($1 {\rm SB} \simeq 0.006$\,MHz). The phase error of these sub bands was estimated at \dtec = 0.4\,TECU. The number of stations ($N$) is taken to be 512.}
    \label{fig:skalow_freqdependence}
\end{figure}

\begin{figure}[t]
    \centering
    \includegraphics[width=\linewidth]{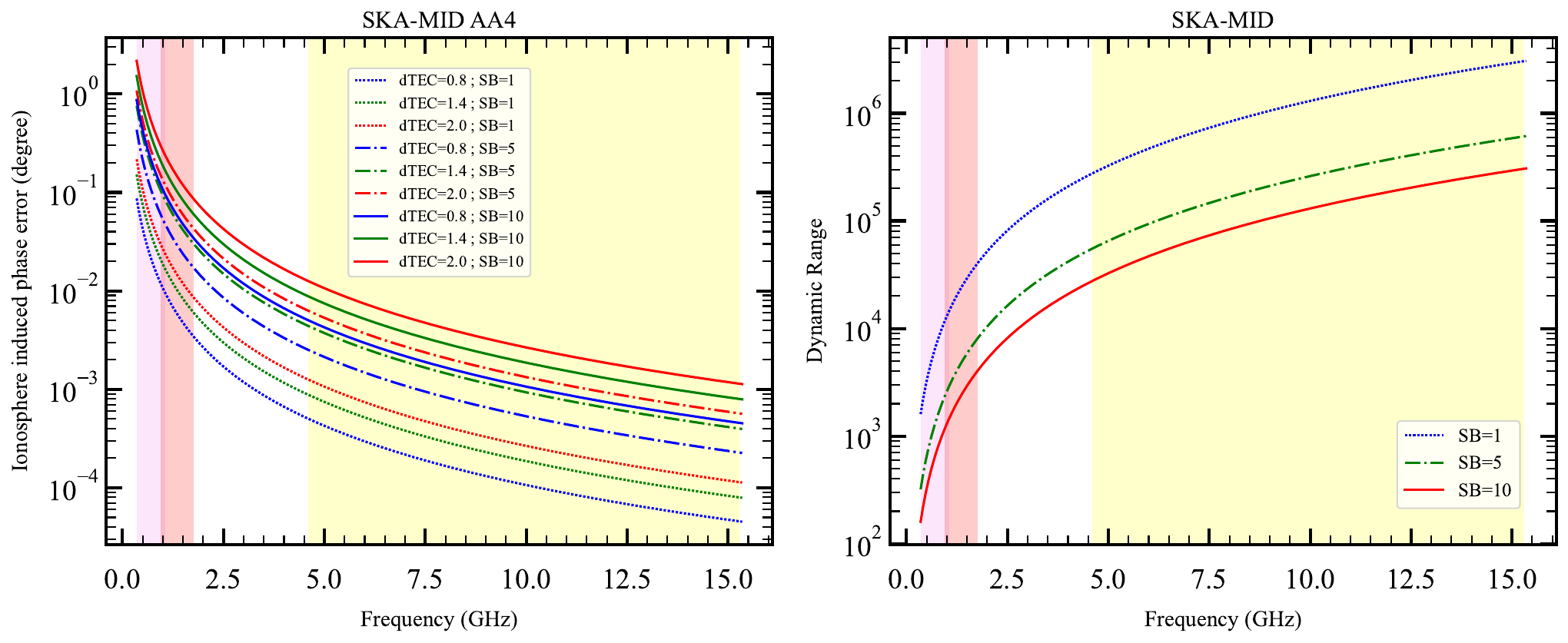}
    \caption{\texttt{Left:} Ionospheric-induced phase variations between the beginning and end of a band of 1, 5, and 10 SKA-Mid AA4 sub bands ($1 {\rm SB} \simeq 0.01375$\,MHz). Assumed \dtec is 0.8, 1.4, 2.0 TECU. SKA-Mid frequency bandwidths are 0.350-1.05\,GHz (Band 1; violet), 0.95-1.76\,GHz (Band 2; red), and 4.6-15.3\,GHz (Band 5; yellow). \texttt{Right: } Dynamic range ($DR$) as a function of frequency for sub bands 1, 5, and 10 ($1 {\rm SB} \simeq 0.01375$\,MHz). The phase error of these sub bands was estimated at \dtec = 0.4\,TECU. The number of antennas ($N$) is taken to be 197.}
    \label{fig:skamid_freqdependence}
\end{figure}

%
When observing a celestial radio source, most low-frequency telescopes work in low signal-to-noise ratio due to low antenna sensitivity and high sky temperature. A typical way to handle this is to average the data in longer time bins or frequency intervals when solving complex antenna gains. This is a trade-off between the signal-to-noise ratio and the decorrelation for finding a good solution. The ionosphere varies rapidly, and averaging over more than 5-10\,sec time typically makes it difficult to trace these variations. Additionally, merging too many frequency channels is not a good idea, as shown in Fig.~\ref{fig:skalow_freqdependence}, between the edges of a five SKA-Low sub-band ($1 {\rm SB} \simeq 0.006$\,MHz) centered around 60\,MHz and considering \dtec value of 1.0\,TECU, there is a differential phase of 10$^\circ$. For core antennas (baselines less than a few kilometers, with \dtec $\lesssim0.4$\,TECU), this constraint can be relaxed, and one can average 10-20 sub-bands easily and still obtain good complex gain solutions. The same is shown for SKA-Mid ($1 {\rm SB} \simeq 0.01375$\,MHz) in Fig.~\ref{fig:skamid_freqdependence}, where averaging over a high number of sub-bands will provide good complex antenna gain solutions as ionospheric induced phase errors are negligible at frequencies above 1\,GHz. \par
%

%
%
In short, for accurate calibration for all baselines and different ionospheric conditions, antenna gain solutions must be calculated at high time and frequency resolution. Also, the left panel of Figures.~\ref{fig:skalow_freqdependence} and \ref{fig:skamid_freqdependence} can be used to estimate the maximum of frequency channels, averaging to compute the complex antenna solutions. \par
\citet{Perley1999ASPC..180..275P} estimated the limitation of dynamic range ($DR$) due to time-independent phase error ($\phi_{err}$ in radians) in all baselines, which will be introduced because of a point source data. The dynamic range of the image is given by the following:
\begin{equation}
    \label{eq:dynamic_range}
    DR \simeq \frac{1}{\phi_{err}}\sqrt{\frac{N(N-1)}{2}}
\end{equation}
\begin{table*}[ht]
\centering
\caption{Typical ionospheric phase errors (in degrees) for SKA-Mid}
\label{tab:ska_mid}
\resizebox{\textwidth}{!}{%
\begin{tabular}{lcccc}
\hline
 & \multicolumn{4}{c}{Frequency} \\
\cline{2-5}
$\Delta$TEC (TECU) & First order & First order & First order & First order \\
                   & 0.77 GHz & 1.4 GHz & 6.7 GHz & 12.5 GHz \\
\hline \hline
2.0 (remote st., bad iono.)  & 1257 & 691 & 144 & 77 \\
1.2 (remote st., good iono.) & 754  & 415 & 87  & 46 \\
0.7 (across FOV)             & 440  & 242 & 51  & 27 \\
\hline
\end{tabular}%
}
\end{table*}
%
where $N$ is the number of interferometer antenna elements. We estimated the $DR$ for SKA-Low and SKA-Mid as a function of frequency for \dtec $\sim$ 0.7 and 1.4\,TECU, respectively, for sub bands (1, 5, and 10). From the right panel of Figures~\ref{fig:skalow_freqdependence} and \ref{fig:skamid_freqdependence}, one can notice that $DR$ increases as a function of frequency, but at the same time, averaging over too many sub bands will decrease it. It is important to note that the $DR$ was estimated in view of a point source in the field. But, by considering the sensitivity of SKA, the $DR$ estimates will be lower than mentioned in the right panel of Figures~\ref{fig:skalow_freqdependence} and \ref{fig:skamid_freqdependence}, because many sources will be present in the FoV. \citet{Datta2009ApJ...703.1851D} showed using simulations that the phase calibration error of 0.1$^{\circ}$ for $N$ = 512 elements array at 158\,MHz will yield a $DR$ of $\sim\,10^{5}$ whereas the desired $DR$ is $\sim\,10^{8}$ to detect the \hi 21-cm signal from reionization.

\subsection{Beam apodization to recover the ionospheric structure }\label{apodization}
For the MWA, ionospheric monitoring is comparatively straightforward at the MRO location because its wide FoV ($\sim 30^{\circ}$) is larger than the characteristic spatial scales of many ionospheric irregularities ($\sim 10 - 100$\,km). At typical ionospheric altitudes of $300 - 400$\,km \citep{jordan2017iono_grad}, these correspond to angular scales of $\sim10^{\circ}$\citep{Helm2020RaSc...5507106H}. In contrast, the planned FoV for SKA-Low will be $3-5$\,deg, depending on the frequency of interest \citep{Braun2019arXiv191212699B}, owing to the larger station size, considerably smaller than the characteristic size of many ionospheric structures. As a result, large-scale coherent ionospheric patterns that move smoothly across the sky within the MWA’s wide FoV may instead appear as more rapid, and seemingly stochastic, direction-dependent phase variations within SKA-Low’s more limited instantaneous view.\\

To effectively probe such large-scale ionospheric structures with SKA-Low, beam apodization \citep{Mort2017MNRAS.465.3680M} and sub-station beam formation \citep{trott_2024_16951143} are expected to play an important role in observing and calibration strategies. Beam apodization\footnote{\href{https://www.skao.int/en/science-users/ska-tools/low-station-beam-simulator}{\url{https://www.skao.int/en/science-users/ska-tools/low-station-beam-simulator}}} applies a spatial weighting function across station elements to shape the primary beam, typically by tapering the outer elements to suppress sidelobes and stabilize the beam response \citep{Peng2024Senso..24.4723P}. This trade-off between sidelobes suppression and modest main beam broadening is well established in array signal processing, where taper functions such as Gaussian, Tukey, Kaiser, Hann and Chebyshev offer different balances between spatial resolution and directivity. Because SKA-Low is a digital interferometric instrument, this technique can be dynamically implemented to modify the effective primary beam shape and to adjust to the desired sensitivity-FoV trade-off. Additionally, selected antennas can be excluded from the station configuration to temporarily reduce the effective station diameter, thereby enlarging the FoV at the cost of sensitivity. This flexibility will allow SKA-Low to better capture and characterize large-scale ionospheric structures that would otherwise appear as small-scale fluctuations in a restricted FoV \citep{Helm2020RaSc...5507106H}. 
%
\subsection{Ionospheric Faraday Rotation and Polarimetric Calibration}
Although dispersive delay dominates the overall ionospheric phase corruption, ionospheric Faraday rotation (FR) introduces an additional frequency-dependent effect that is particularly important for polarimetric observations. As a linearly polarised radio wave propagates through the magnetised ionospheric plasma, the plane of polarisation undergoes a rotation given by

\begin{equation}
\Delta\Omega_{\mathrm{FR}} = \mathrm{RM}_{\mathrm{ion}} \lambda^{2}
\end{equation}
where the ionospheric rotation measure is defined as
\begin{equation}
\mathrm{RM}_{\mathrm{ion}} = \frac{e^{3}}{8\pi^{2}\varepsilon_{0}m_{e}^{2}c^{3}} \int n_{e}(l) B_{\parallel}(l) \, \mathrm{d}l.
\end{equation}

Here, $n_{e}$ is the electron density and $B_{\parallel}=B\cos\theta$ is the magnetic-field component along the line of sight. Because $\Delta\Omega_{\mathrm{FR}} \propto \lambda^{2}$, the effect dominates below $\sim 80$ MHz and on baselines $\gtrsim 10$ km, where spatial variations in $n_{e}$ and $B$ produce differential FR between stations \citep{Mevius2018a}. Ionospheric FR has several observationally significant effects on polarized emission beyond the scalar amplitude and phase corruptions discussed above. The most direct effect is a frequency- and time-dependent rotation of the observed electric vector position angle\,(EVPA). As the EVPA rotates, it continuously mixes the linear Stokes parameters $Q$ and $U$ \citep{Hamaker1996A&AS..117..137H}. If this rotation is left uncorrected, averaging the data over a finite frequency channel\,(bandwidth depolarization) or a finite integration time\,(time smearing) will artificially wash out the intrinsic polarized signal of the astrophysical source \citep{Burn1966, Sokoloff1998}. Furthermore, uncorrected differential FR across the array introduces amplitude and phase decorrelation of the observed polarized visibilities \citep{Perley2026ApJS..283...82P}. In combination with instrumental polarisation, these effects can exacerbate polarisation leakage caused by imperfect calibration and distort RM-synthesis spectra \citep{brentjens05}. Even for the sensitive 21-cm cosmology experiments, these effects are particularly problematic because spectrally varying polarized foregrounds can leak into Stokes $I$ via instrumental polarization, introducing strong frequency-dependent structure, which can mimic or obscure the cosmological signal \citep{jelic10, moore13, nunhokee17, spinelli19}.

Achieving high-fidelity polarimetry, therefore, requires accurate correction of the ionospheric rotation measure. For current low-frequency radio telescopes, calibration to sub-rad\,m$^{-2}$ precision is necessary. LOFAR observations of bright pulsars have demonstrated that, after removing diurnal trends, ionospheric RM corrections can reach an accuracy of $\sim 0.06$–$0.07$\,rad\,m$^{-2}$ \citep{Porayko2019MNRAS.483.4100P}. To achieve this, tools such as the publicly available code \texttt{ionFR}\footnote{\href{https://github.com/csobey/ionFR}{\url{https://github.com/csobey/ionFR}}} \citep{Sotomayor-Beltran2013A&A...552A..58S} and \texttt{RMextract} \citep{Mevius2018ascl.soft06024M} compute $\mathrm{RM}_{\mathrm{ion}}$ using GPS-derived total electron content maps in combination with geomagnetic field models (e.g., IGRF), enabling the precision required for pulsar RM and Galactic magnetism studies. While \texttt{ionFR} is no longer actively maintained, \texttt{RMextract} has recently been superseded by \texttt{spinifex}\footnote{\href{https://git.astron.nl/RD/spinifex}{\url{https://git.astron.nl/RD/spinifex}}} \citep{mevius_2025_spinifex}, which is under development for next-generation low-frequency arrays such as LOFAR and SKA-Low. At higher frequencies, interferometers such as the VLA and MeerKAT require RM accuracies of $\lesssim 0.1$\,rad\,m$^{-2}$ at metre wavelengths to constrain the EVPA to within $\sim 5^{\circ}$ \citep{Perley2026ApJS..283...82P}. Even after applying TEC-based corrections, residual ionospheric RM fluctuations at the level of $\sim0.1$--$0.5$\,rad\,m$^{-2}$ can persist on short timescales \citep{Porayko2019MNRAS.483.4100P, gasp18}; these can be further mitigated by self-calibration strategies that solve for direction-dependent RM corrections using sufficiently bright polarised calibrators within the field of view \citep{Intema2009A&A...501.1185I}.

In the SKA era, these lessons will be combined. \texttt{spinifex}-type modeling will provide the deterministic correction for wide-field polarization imaging. However, stochastic temporal variability -- identified in HERA 21\,cm experiments as a depolarization mechanism -- must be treated statistically to avoid biasing deep cosmological measurements \citep{Martinot2018ApJ...869...79M}. This unified approach, bridging deterministic calibration and statistical mitigation, will preserve polarimetric fidelity for both SKA-Low and SKA-Mid science cases. Ignoring this effect would undermine key SKA science objectives, including studies of cosmic magnetism, pulsar emission, and tests of fundamental physics.

\subsection{Implications for SKA}
The diverse ionospheric techniques developed on precursor and pathfinder instruments worldwide for SKA - antenna-based phase/TEC tracking, field-based astrometric (positional-shift) measurements, and amplitude scintillation analysis - are expected to converge in the SKA era, providing a more comprehensive and self-consistent view of ionospheric structure and dynamics. By overcoming current geographical and instrumental limitations, the complementary capabilities of SKA-Low and SKA-Mid will enable ionospheric science across a wide range of scales, spanning spatial scales from sub-kilometre plasma irregularities to global space-weather dynamics.

\paragraph{Micro-scale precision: resolving small-scale phase structure.}
At the smallest scales, the SKA-Low, with its unprecedented sensitivity, frequency coverage, large collecting area, and compact core configuration, will significantly improve upon antenna-based differential TEC measurements with precision significantly below the $10^{-3}\,\mathrm{TECU}$ level achieved with current instruments such as uGMRT, LOFAR and VLA. The dense sampling of baselines will enable high-fidelity modeling of the TEC surface across the array (e.g., third- and fourth-order polynomial fitting), thereby enabling the reconstruction of increasingly complex, non-linear ionospheric structures. This improved sensitivity and baseline density may allow access to sub-kilometre plasma fluctuations. Recent empirical studies utilizing MWA Phase-II extended baselines demonstrate that non-linear structures larger than $0.6$\,km aggressively dominate the phase screen at baselines extending beyond $2$\,km \citep{Rioja2022}. Combined with advanced direction-dependent calibration algorithms such as Low frequency Excision of Atmosphere in Parallel (LEAP), these measurements provide a pathway toward more realistic three-dimensional defocusing effects, pushing low-frequency interferometry beyond the flat-screen approximation \citep{Rioja2018, Rioja2022}.

\paragraph{Macro-scale dynamics: wide-field mapping and TIDs.}
On larger spatial scales, the wide field of view of SKA-Low allows field-based calibration using the expected density of compact background sources, about $3\times10^{4}$ in number \cite[see Table 3 of][]{Braun2019arXiv191212699B}, exceeding 1.5 mJy in apparent flux density (leveraging catalogues like TRECS; \citealt{bonaldi2019}). This generates a highly dense, isotropic grid of ionospheric pierce points across the primary beam. Instead of merely measuring linear gradients, the SKA will map non-linear ionospheric distortions (e.g., curvature and wave-like structure) with high cadence. Such capability is well-suited to tracking the propagation of TIDs and related wave-like structures, including medium-scale TIDs potentially associated with sporadic-E layers. In this sense, SKA-Low can function as a high-fidelity, wide-field ionospheric monitor. In parallel, SKA-Mid, with its multi-band receivers and longer baselines, will enhance sensitivity to small residual phase errors and enable cross-frequency validation through the expected $\nu^{-1}$ scaling of ionospheric phase shifts in the cold-plasma limit. 

\paragraph{Amplitude Scintillation: beyond phase delays.}
The SKA will probe fine-scale plasma irregularities through amplitude scintillation. The broad frequency coverage of SKA-Low is expected to provide sensitivity to multi-layer scattering effects, potentially revealing deviations from the standard thin-screen approximation. Furthermore, MWA measurements have shown that amplitude scintillation probing Fresnel scales of $\sim300$\,m is highly correlated with larger-scale refractive shifts \citep{Waszewski2022PASA...39...36W}. This correlation means SKA-Low will be able to construct a unified empirical model of the ionospheric turbulence cascade, linking macroscopic wave generation directly to sub-kilometre dissipation scales. Therefore, combined high-resolution TEC and scintillation maps estimated from SKA observations have the potential to provide new constraints for space-weather forecasting. When incorporated into larger-scale models, these datasets will significantly improve characterization of ionospheric disturbances (e.g., severe geomagnetic storms and plasma bubble dynamics) relevant to navigation and communication systems.

\paragraph{Magnetospheric Coupling and Fundamental Physics:}
Finally, the combination of precise TEC estimation with polarimetric calibration opens new avenues for fundamental physics. While refractive shifts constrain the electron content, measurements of differential Faraday rotation provide sensitivity to the line-of-sight magnetic field weighted by electron density. To achieve this, the SKA will utilize a dense grid of bright, polarized pulsars to track FR variations across the sky, effectively separating cosmic magnetism from atmospheric corruption with the requisite sub-rad m$^{-2}$ precision \citep{Sotomayor-Beltran2013A&A...552A..58S}. As demonstrated by tools such as \texttt{spinifex} \citep{mevius_2025_spinifex} and uGMRT observations \cite[e.g.,][]{mangla2023spectral}, accurate calibration of this effect is essential for separating astrophysical signals from ionospheric contamination. Furthermore, the wide-band capabilities of SKA-Mid will allow us to test the frequency-dependent validation of the theoretical scaling of ionospheric TEC and FR effects, providing a valuable experimental platform for ionospheric plasma physics. 

\section{Conclusion}\label{sec:summary}
The effects of the ionosphere present a primary limiting factor on the quality of low-frequency radio interferometric observations. The ionosphere's ionized plasma has a refractive index that changes over time and space. This variability causes a dispersive delay in the signal, which is dependent on the direction of observation. Ultimately, this directional-dependent delay directly impacts the visibility phase measured by any radio interferometer. The SKA's unique geophysical location can be used to better comprehend and characterise the ionosphere. With the SKA now in its construction phase, there is a timely opportunity to advance calibration strategies and data analysis techniques aimed at detecting and correcting ionospheric phase errors. These approaches will ultimately enable precise mitigation of ionospheric effects from low-frequency interferometric data. SKA-Low's unprecedented sensitivity and wide bandwidth at low frequencies will transform ionospheric scintillation from an observational nuisance requiring data rejection into a rich scientific opportunity for space weather science. SKA-Mid multi-frequency observations will provide the frequency dependence necessary for theoretical interpretation.


\section*{Author contributions}
Abhirup Datta led the conceptual development of the chapter and coordinated the overall writing process. The other authors made significant contributions to the writing, as well as to the structuring, refinement, and final polishing of the chapter. 

\bibliographystyle{abbrvnat}
\bibliography{chapter} 

\end{document}

%% file: authors.tex
\author[1,*]{Abhirup Datta\orcidlink{0000-0002-5333-1095}}\emailAdd{abhirup.datta@iiti.ac.in}
\author[1]{Samit Kumar Pal\orcidlink{0000-0002-2271-4165}}
\author[2]{Sarvesh Mangla\orcidlink{0000-0003-3485-5122}}
\author[1]{Bhuvnesh Brawar\orcidlink{0009-0000-1529-3270}}
\author[3]{Sumanjit Chakraborty\orcidlink{0000-0001-8455-9722}}
\author[4]{Sudipta Sasmal\orcidlink{0000-0003-2703-8861}} 
\author[1]{Anshuman Tripathi\orcidlink{0000-0002-5091-9950}}

\affiliation[1]{Department of Astronomy, Astrophysics and Space Engineering, Indian Institute of Technology Indore, Indore 453552, India}
\affiliation[2]{University Observatory, Faculty of Physics,  Ludwig-Maximilians-Universität, Scheinerstr. 1, 81679, Munich, Germany}
\affiliation[3]{Indian Institute of Geomagnetism, Navi Mumbai, India}

\affiliation[4]{Space and Atmospheric Science, Institute of Astronomy Space and Earth Science, Kolkata, 700054, India}